\documentstyle[sprocl,epsfig]{article}
\def\Journal#1#2#3#4{{#1} {\bf #2}, #3 (#4)}

\newcommand{\half}{{\textstyle{\frac{1}{2}}}}

\newcommand{\fin}{\hspace{0.6cm}}

\newcommand{\trc}{{\rm Tr\hspace{0.2ex}}}

\newcommand{\bgeq}{\begin{equation}}
\newcommand{\bgeqa}{\begin{eqnarray}}
\newcommand{\edeq}{\end{equation}}
\newcommand{\edeqa}{\end{eqnarray}}

\newcommand{\ainv}{a^{-1}}

\newcommand{\ectv}[1]{\left\langle #1 \right\rangle}

\title{SIMULATIONS OF THE ELECTROWEAK PLASMA AT FINITE TEMPERATURE
\footnote{Talk presented at ``Strong and Electroweak Matter '97'', 
21 -- 25 May '97, Eger, Hungary}
}
\author{Joachim Hein%
}
\address{Dept.~Physics \& Astronomy, University of Glasgow, Glasgow G12 8QQ, UK}
\begin{document}
\vspace*{-2cm}
\begin{flushright}
\textbf{\textsf{GUTPA/97/8/1}}\\
\textbf{\textsf{hep-lat/9708001}}\\[2ex]
\end{flushright}
\maketitle
\abstracts{We investigate the SU(2) Higgs model at a Higgs Boson mass
of $\simeq 34$~GeV for temperatures at the electroweak scale. We discuss in
detail the critical temperature, 
the scalar field vacuum expectation
value and the latent heat. We also consider for which
temperatures the plasma can be regarded as radiation dominated.
}

\section{Introduction}
In the standard electroweak theory, the baryon violating processes
are suppressed for temperatures below the electroweak scale
($\approx{}100$~GeV). Therefore the currently observed baryon asymmetry was
finally determined at this temperature scale. Since for these
temperatures infrared
singularities render perturbation theory 
uncertain~\cite{perturb}$^{\!,\,}$\cite{bufohe}, 
a non-perturbative treatment
of the plasma is needed, which can be provided by numerical lattice 
simulations. 
For this purpose we performed large scale lattice simulations of the
SU(2) Higgs model in 4 Euclidean dimensions. Our
results~\cite{first}$^{\!-\,}$\cite{ztalk} also provide an
estimation of the reliability~\cite{34vergl} of the reduction step,
which is used to relate simulation results obtained in effective
models in three dimensions~\cite{3dim}$^{\!,\,}$\cite{3dtalk} to physics.

The results detailed in this talk are obtained with a Higgs boson mass of
$M_H \simeq 34$~GeV, for which a strong first order
phase transition is observed at the electroweak scale. In the
simulations we use the
following action with inverse
lattice spacings in the range $2T_c \le \ainv \le 5T_c$
\bgeqa\label{action}
S[U,\varphi] &=& \beta \sum_{pl}
\left( 1 - \half {\rm Tr\,} U_{pl} \right)
+ \sum_x \bigg\{ \half{\rm Tr\,}(\varphi_x^+\varphi_x) \nonumber\\
&&\fin +
\lambda \left[ \half{\rm Tr\,}(\varphi_x^+\varphi_x) - 1
\right]^2 - \kappa\sum_{\mu=1}^4
{\rm Tr\,}(\varphi^+_{x+\hat{\mu}}U_{x\mu}\varphi_x)
\bigg\} \,.
\edeqa
Here $U_{x,\mu}$ denotes the gauge link variable, $U_{pl}$ the
smallest Wilson loop and $\varphi_x$ is the scalar field in
$2\!\otimes\! 2$ isospin matrix notation. 

\section{Lines of constant physics and critical temperature}
To keep the renormalised couplings $g_R$ and $\lambda_R$ constant when
stepping down in the lattice spacing $a$, we guess the bare $\beta$
and $\lambda$ from the 1-loop renormalisation group equation. 
The hopping parameter $\kappa$ is tuned  
to its critical value in a simulation
at a temporal lattice extent $L_t$, as detailed 
in Ref.~\cite{plasmapap}. 
Which determines the critical temperature to be $aT_c = L_t^{-1}$.

\begin{figure}[t]
\centerline{
\epsfig{file=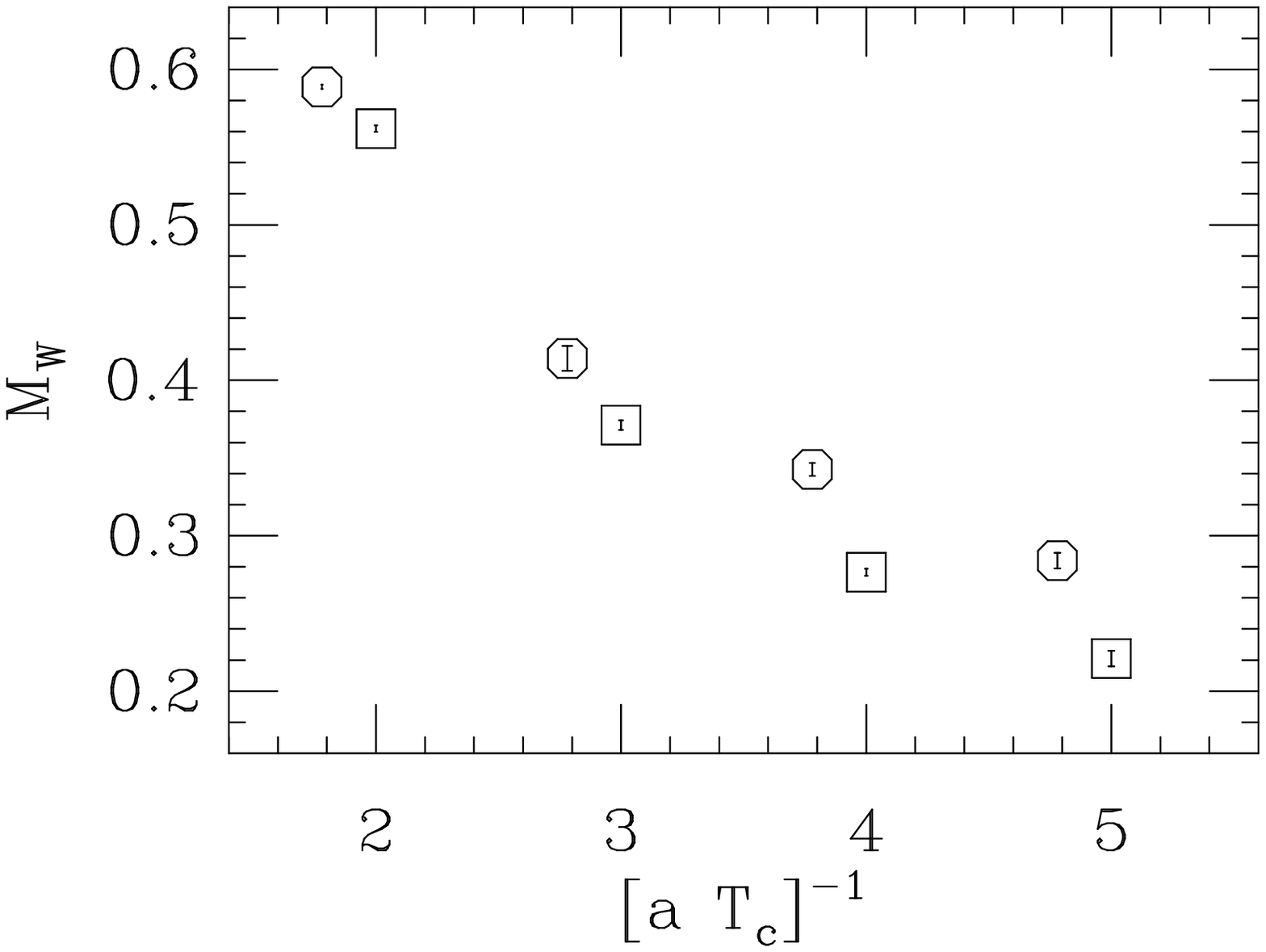,height=2.75cm,
bbllx=49pt,bblly=35pt,bburx=525pt,bbury=385pt}
\epsfig{file=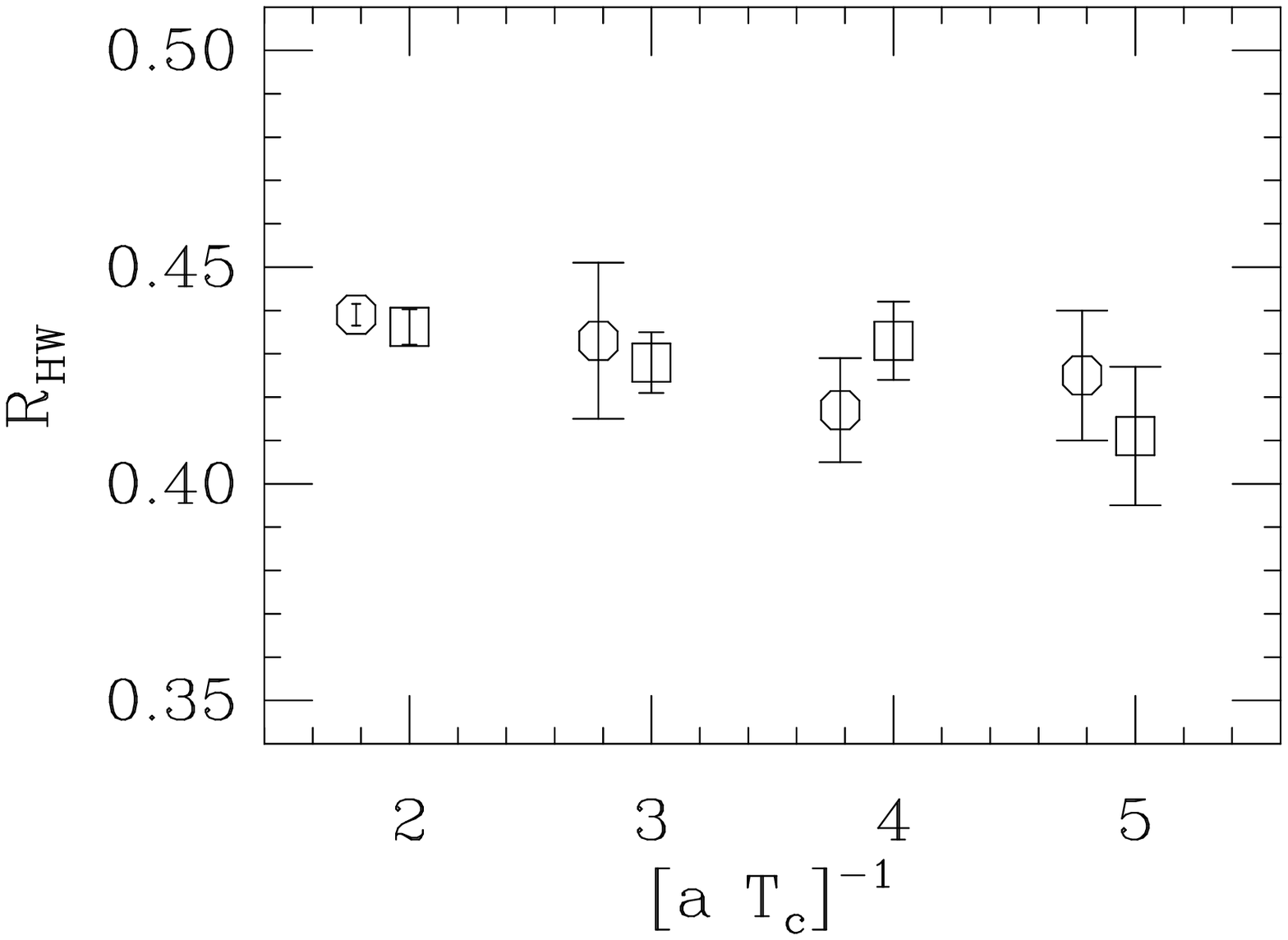,height=2.75cm,
bbllx=34pt,bblly=35pt,bburx=525pt,bbury=385pt}
\epsfig{file=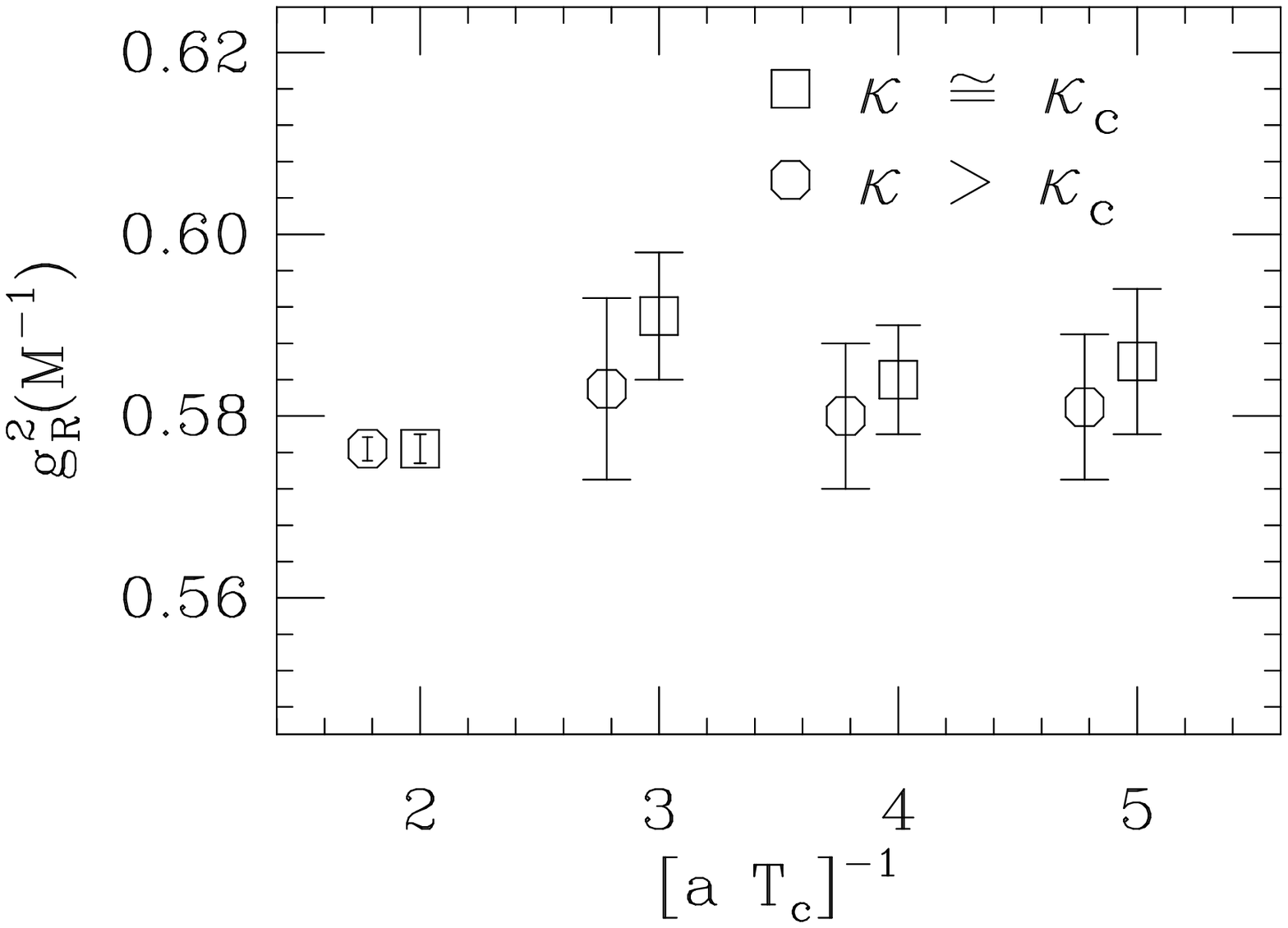,height=2.75cm,
bbllx=27pt,bblly=35pt,bburx=525pt,bbury=385pt}}
\caption{\label{renaflow}
Flow of the renormalised parameters with the lattice spacing.
For $R_{HW}:=M_H/M_W$ we have $g_R R_{HW}= (32 \lambda_R)^{1/2}$.
The octagons
refer to points with $\kappa$ shifted by $5\cdot 10^{-5}$ resp.\
$4\cdot 10^{-5}$ in case of the finest lattice, when compared to the
squares.}
\end{figure}

In order to obtain the size of the lattice spacing in physical units,  
the  Higgs and W-boson
masses and the renormalised coupling~\cite{first} are measured at
$T=0$. In a detailed finite volume study at $\ainv = 2T_c$ we find 
a spatial lattice extent $L_s \ge 12$, which relates to the correlation
length of the Higgs boson as $L_s \ge 2.9\, \xi_H$, to be sufficient to 
keep the renormalised
parameters constant within an accuracy of 1\% or even
better~\cite{plasmapap}.
Using these physical volumes we investigate the influence of the
finite $a$ and the uncertainty of the critical hopping parameter on
the renormalised parameters.
The outcome 
is shown in figure \ref{renaflow}, where 
a small shift of $\kappa$ affects only the W-mass in lattice units.
This leads to a
significant contribution to the error of $a$ from the
uncertainty of $\kappa$.  
The renormalised couplings $\lambda_R$ and
$g_R$ however are unaffected within their small errors.
A second important result is that the renormalised
couplings are unchanged when the lattice spacing is decreased.
This means that our simulations indeed follow the lines of constant physics.

\begin{figure}[t]
\centerline{\epsfig{file=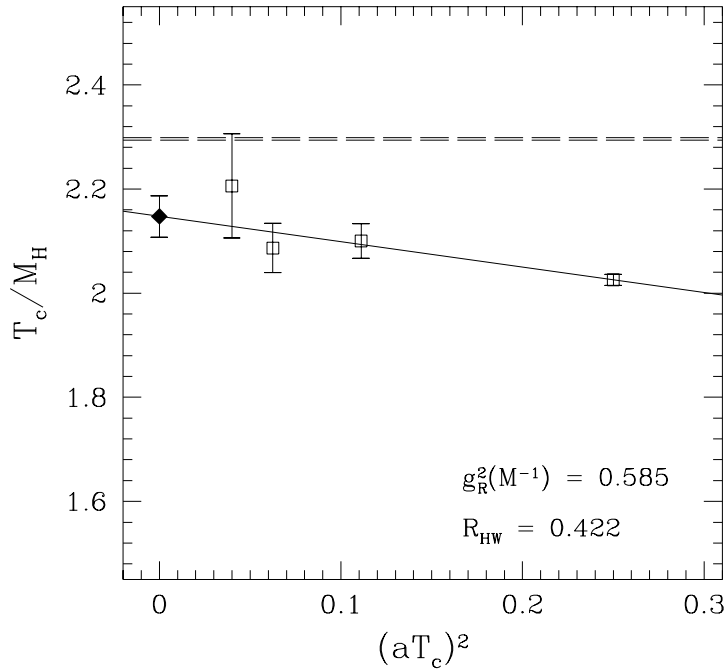,width=9.5cm}}
\caption{\label{tcmhfig} Extrapolation of $T_c/M_H$ to the continuum
limit which is given by the filled symbol. 
The dashed lines represents the perturbative
predictions~\protect\cite{bufohe}. }
\end{figure}

We are now in the position to give a precise
estimate of the critical temperature in physical units. This is shown
in figure~\ref{tcmhfig}. 
The results from the previous paragraph are
given by the white symbols. Their error bars contain the
uncertainties of $\kappa_c$ and of both of the $M_H$ values.  Since the
lattice artifacts of eq.~(\ref{action}) are ${\cal O}(a^2)$, we
extrapolate with a quadratic ansatz to small $a$-values, 
which is justified from the good $\chi^2 \simeq 1$.
The result $T_c/M_H|_{a\to 0} = 2.15(4)$ is shown by the
filled symbol. Note the surprisingly small scaling
violations of only 5\%, when comparing the result at $\ainv = 2T_c$ to
the extrapolated value. 
With $M_W \simeq 80$~GeV we get $T_c=72.8\pm 1.3$~GeV for $a\to 0$.

The perturbative estimate~\cite{bufohe} shown by the
dashed line, differs from the extrapolated lattice 
result by three standard deviations, so one cannot exclude non-negligible 
higher order or non-perturbative contributions.

\section{Scalar field vacuum expectation value}
\begin{figure}[t]
\centerline{\epsfig{file=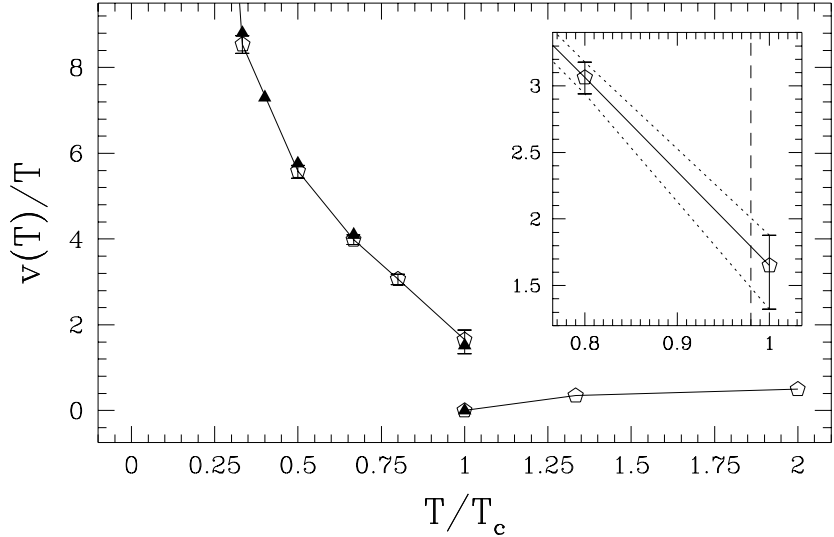,width=11cm}}
\caption{\label{vtfig} Result for $v/T$ as a function of temperature. Filled
triangles denote results for $\ainv = 2T_c$, the pentagons those for
$4T_c$. The error bars in the enlargment give 
the uncertainty of $\kappa_c$ only. The vertical dashed
line gives the thin wall estimate of the
supercooling~\protect\cite{jhdiss}.} \end{figure}

The scalar field vacuum expectation value $v$ plays a prominent r\^ole
in the semi-classical estimation of the sphaleron rate in the symmetry
broken phase. A gauge invariant renormalised vacuum expectation value
is defined by
$v_R := 2 M_W/g_R$.
However since the mass determination is demanding in terms of CPU time, 
we approximate this by
$v(T) := [2\kappa\left(\rho^2_x(T)-\rho^2_{x,{\rm
sym}}(T_c)\right)]^{1/2}$ with 
$\rho_x^2 := \half \trc (\varphi^+_x \varphi_x)$.
It has been checked that $v_R$ and $v(T)$ are in reasonable agreement for 
$T=0$ and $T \approx T_c$~\cite{jhdiss}. 
The result for $v/T$ is given in figure~\ref{vtfig}.
We also given an estimate of
the supercooling obtained in the thin wall approximation. 
From this one can expect at most an increase of 15\% in the
exponent of the sphaleron rate due to supercooling. 
With the Clausius-Clapeyron Equation~\cite{bufohe}
the latent heat can be estimated from the discontinuity at $T_c$.
\bgeq \label{cclheat}
\Delta \epsilon/T_c^4 = 0.281(19)|_{\ainv=2T_c}\,,\fin
\Delta \epsilon/T_c^4 = 0.31(12)|_{\ainv=4T_c}\,.
\edeq

\section{Thermodynamics of the plasma}
\begin{figure}[t]
\centerline{\epsfig{file=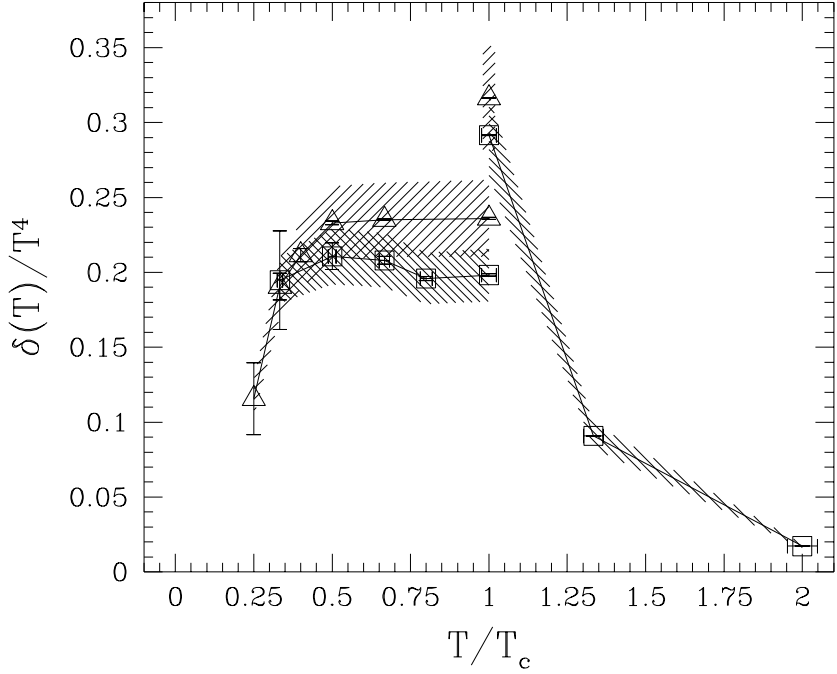,
width=11cm}}
\caption{\label{deltafig} Result for $(\frac{1}{3}\epsilon-P)T^{-4}$.
Triangles represent $\ainv = 2T_c$, squares $4T_c$.}
\end{figure}

In this section we discuss the thermodynamic quantity $\delta :=
\frac{1}{3}\epsilon-P$ which measures the deviation of
the plasma from pure radiation.
With $\epsilon$ the energy density and with $P$ the pressure is denoted,
while $\delta$ can be determined from
\bgeq \label{deltaeq}
\frac{\delta}{T^4} = 
\frac{(L_t)^4}{3} \left.\left[
8\frac{\partial \kappa}{\partial \tau} 
   \ectv{L_\varphi}
-\frac{\partial \lambda}{\partial \tau} \ectv{(\rho^2_x \!-\!1 )^2}
-6\frac{\partial \beta}{\partial \tau} \ectv{1\!-\!\half \trc U_{pl}}
\right]\right|_{g_R, \lambda_R} \,,
\edeq
with $\tau := -\log(M_W)$ and
$L_\varphi := \half\trc (\varphi^+_{x+\hat \mu} U_{x,\mu}\varphi_x)$.
We determine the derivatives of $\beta$ and
$\lambda$ using the 1-loop renormalisation group
equations and those of $\kappa$ from fits to the simulation
results for $\kappa_c$~\cite{plasmapap}. 
Since eq.~(\ref{deltaeq}) contains divergent
vacuum contributions, $\delta(T\!=\!0)$ must 
be subtracted to obtain the physical
result, which is show in figure~\ref{deltafig}. 
The 
figure shows different errors: statistical errors by
vertical bars, uncertainty of $\kappa_c$ by horizontal bars and the
uncertainty arising from $\frac{\partial \kappa}{\partial \tau}$ by
the shaded region. 
In the symmetric phase $\delta/T$ goes down by one order of magnitude
when $T$ is increased from $T_c$ to $2T_c$ where it is almost
compatible with 0. This strong decay is
confirmed by the good agreement of the slopes $\frac{\partial\delta/T
}{\partial \kappa}$ for both $a$-values~\cite{plasmapap}, so that
for $T\ge2T_c$ the plasma can be considered to be radiation dominated.

Again  from the jump at $T_c$ the latent heat can be determined:
\bgeq
\Delta \epsilon/T_c^4 = 0.240(30+4)|_{a^{-1}=2T_c}\,,\fin
\Delta \epsilon/T_c^4 = 0.28(3+9)|_{a^{-1}=4T_c}\,.
\edeq
The first number in the brackets represents the statistical error and
the second one the uncertainty of $\kappa_c$. The result is in
agreement to eq.~(\ref{cclheat}) and no scaling violation is to be
observed between the results at $\ainv = 2 T_c$ and $4T_c$.

\section*{Acknowledgements}
I thank F.~Csikor, Z.~Fodor, J.~Heitger, K.~Jansen,
A.~Jaster and I.~Montvay warmly for collaboration.
The simulations have been performed on computers of DESY-IfH and HLRZ.
Financial support by the  European commision under Ref.~No.:~ERB FMB ICT 
961729 is gratefully acknowledged.


\begin{thebibliography}{99}
\bibitem{perturb}P.~Arnold, O.~Espinosa, \Journal{Phys.\ Rev.\
D}{47}{3546}{1993}; \Journal{Erratum-ibid}{50}{6662}{1994}; 
W.\ Buchm\"uller et al., \Journal{Ann.\
Phys.}{234}{260}{1994}; Z.~Fodor, A.~Hebecker, \Journal{Nucl.\ Phys.\
B}{432}{127}{1994}
\bibitem{bufohe}W.\ Buchm\"uller, Z.~Fodor, A.~Hebecker, \Journal{Nucl.\
Phys.\ B}{447}{317}{1995}
\bibitem{first}
Z.~Fodor, et al.,
\Journal{Phys.\ Lett.\ B}{334} {405}{1994};
\Journal{Nucl.\ Phys.\ B}{439} {147} {1995}
\bibitem{interface}F.~Csikor, Z.~Fodor, J.~Hein, J.~Heitger,
  \Journal{Phys.\ Lett.\ B} {357} {156}{1995}
\bibitem{plasmapap}
F.~Csikor, et al.,
\Journal{Nucl.\ Phys.\ B}{474}{421}{1996}
\bibitem{jhjh}J.~Hein, J.~Heitger, 
 \Journal{Phys.\ Lett.\ B} {385} {242}{1996}
\bibitem{jhdiss}J.~Hein, Doctoral thesis (1996), Universit\"at Hamburg 
(German),\newline
[\texttt{http://www.physics.gla.ac.uk/}\verb+~+\texttt{jhein/}]
\bibitem{ztalk}{Talk presented by Z.~Fodor, these proceedings}
\bibitem{34vergl}M.\ Laine, \Journal{Phys.\ Lett.\ B}{385}{249}{1996}
\bibitem{3dim} see e.g.: K.~Kajanite et.al., 
\Journal{Nucl.\ Phys.\ B}{466} {189} {1996}; 
\Journal{Phys.\ Rev.\ Lett.}{77}{1996}{2887}; 
O.~Philipsen, et.\ al., \Journal{Nucl.\ Phys.\ B}{469}{445}{1996};
F.~Karsch, et.\ al., \Journal{Nucl.\ Phys.\ B (Proc.\ Suppl.)}{53}{417}{1997};
M.~G\"urtler, et.\ al.,
\Journal{Nucl.\ Phys.\ B}{483} {383} {1997}; UL-Preprint (1997) {\tt
hep-lat/9704013v3}
\bibitem{3dtalk}
Talks presented by M.~Laine and O.~Philipsen, these proceedings
\end{thebibliography}
\end{document}